\documentclass[11pt]{article}
\usepackage[T1]{fontenc}
\usepackage[utf8]{inputenc}
\usepackage[a4paper,margin=1in]{geometry}
\usepackage{amsmath}
\usepackage[numbers,sort&compress]{natbib}
\usepackage{url}
\usepackage{graphicx}
\usepackage{color}
\usepackage{microtype}
\usepackage{tikz}
\usetikzlibrary{arrows.meta,positioning,fit}
\usepackage[hidelinks]{hyperref}

\title{Automatic Transcription of Microtonal Free-Rhythm Vocal Music: A Case Study in Iranian Classical Music}
\author{Sepideh Shafiei\thanks{Corresponding author: \href{mailto:sepideh.shafiee@gmail.com}{sepideh.shafiee@gmail.com}} \quad Shapour Hakam \quad Harsh Dange \quad Joel Rodriguez Caraballo\\[0.5em]
\normalsize Cu Test Inc., Berkeley, California, USA}
\date{}

\hypersetup{
  pdftitle={Automatic Transcription of Microtonal Free-Rhythm Vocal Music: A Case Study in Iranian Classical Music},
  pdfauthor={Sepideh Shafiei, Shapour Hakam, Harsh Dange, and Joel Rodriguez Caraballo}
}

\begin{document}

\maketitle

\begin{abstract}
This paper introduces a computational workflow for automatically transcribing microtonal, free-rhythm vocal music, with Iranian classical music as a case study. Our approach is based on performances by the renowned vocalist Karimi and ground truth transcriptions by the prominent ethnomusicologist Masoudieh \cite{M1}, which were subsequently incorporated into the IRMA Audio-MIDI dataset \cite{IRMA2025}. To accurately extract melodies, we employ pitch histograms in conjunction with Dynamic Time Warping (DTW). Additionally, we introduce specialized musical notations to capture the intricate ornamentations characteristic of the genre, with particular emphasis on the vocal technique \emph{tahrir}. The transcription process is implemented in Python using the \emph{music21} library for symbolic music representation \cite {Cuthbert}. This study not only advances the field of computational ethnomusicology but also highlights the potential of computational methods in preserving and analyzing complex musical traditions. The transcription system also generates a combined visualization of the audio pitch contour and the DTW-aligned MIDI representation, enabling users to inspect the correspondence between the performance and the generated transcription. A companion visual editor supports expert-in-the-loop correction of the resulting notation.

\end{abstract}
	
	\section{Introduction}\label{sec:introduction}
	Existing approaches to Automatic Music Transcription (AMT) have largely focused on tonal, metrical music, where established signal processing methods, probabilistic models, NMF, and neural networks have been employed to achieve reasonable transcription accuracy \cite{Benetos, Martak}. 
	However, the existing methods often fall short when applied to non-Western microtonal or free-rhythm music, especially in vocal genres where expressive deviations in pitch and rhythm are an essential feature. Additionally, specialized musical notation systems, such as those used to transcribe ornamentations like \emph{tahrir}, are essential for accurately representing the melodic subtleties of Iranian classical music. 
	These systems must account for microtonal intervals that fall between Western pitches, as well as the complex pitch modulations characteristic of the style. 
	There are few studies on automatic transcription of non-Western music among which one can mention Turkish Makam Music \cite{Benetos2}, \cite{Gedik}, \cite{Germen}, Arabic and Turkish music \cite{wb}, and Hindustani vocals \cite{Jain}.
	Our transcription software is tailored for Iranian classical music, which features non-standard intervals \cite{S-11}, complex ornamentations \cite{S-12}, and diverse modes that are difficult to capture with Western transcription systems. The free-rhythm nature and absence of a fixed meter in this music, especially in vocal performances, demand advanced computational techniques for accurate representation.

By utilizing pitch histograms in multiple levels in conjunction with DTW for melody extraction, and leveraging the \emph{music21} library for symbolic music representation, we develop a transcription system that captures the intricate ornamentation and variable tuning of this genre.  Using optimization methods we approximate the tempi of the pieces. 

Developed in collaboration with musicians, the proposed framework consists of two complementary programs. The first performs automatic transcription and uses DTW to align the audio pitch contour with the generated MIDI representation, displaying both in a single graph for inspection. The second is a companion visual editor that enables musicians and music engravers to review and correct the generated transcription interactively. Together, these programs combine automated processing with expert supervision while preserving the musical and cultural nuances of the repertoire. The framework is also adaptable for use in related musical traditions, particularly in free-rhythm and modal performances. This approach not only advances the field of automatic music transcription (AMT) but also highlights the importance of tailoring transcription models to specific musical and cultural contexts. 

Section \ref{names} provides an overview of all the Iranian musical terms and concepts mentioned throughout the text, arranged in alphabetical order, to assist readers unfamiliar with Iranian music. 


\section{Core Components of the Transcription Process}
\label{Methodology}

\begin{figure}[t]
\centerline{
\includegraphics[width=0.9\linewidth]{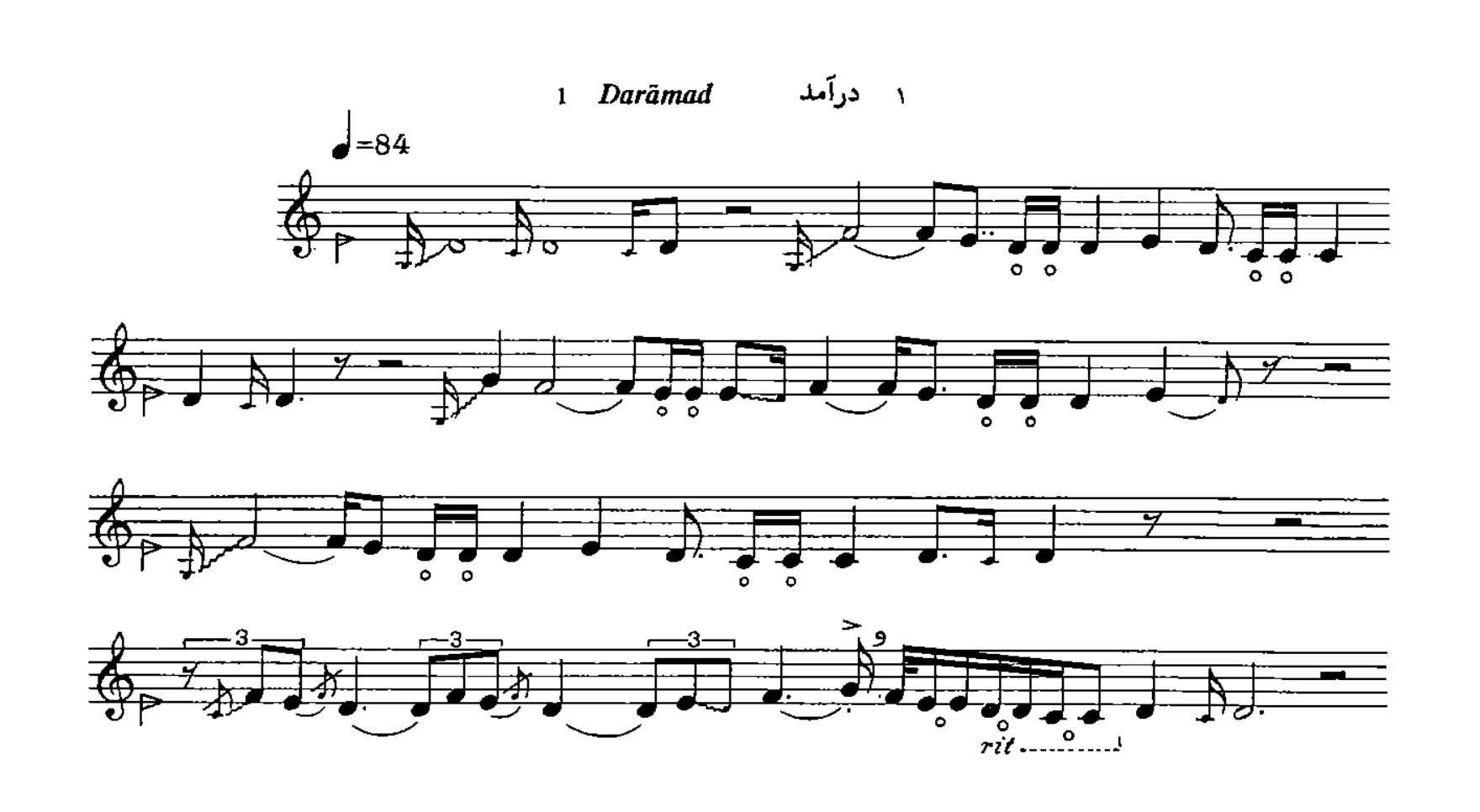}}
 \caption{Masoudieh's Transcript of Example 1}
 \label{daramad}
\end{figure}

In this section, we outline the core components of the data preprocessing, along with the essential musical and computational background. To clarify the process, we focus on a specific example from Karimi's repertoire, referred to as Example 1 throughout the paper. Figure \ref{daramad} presents Masoudieh's transcription of the piece, which serves as our reference \cite {M1}. The small hollow circles placed above or below certain notes indicate \emph{tekye}, a rapid transition in tahrir from the main note to a higher-pitched note. The p-shaped sign associated with the note E, known as \emph{koron}, denotes a pitch that is approximately half-flat. However, the precise intonation is a topic of debate among musicians and musicologists, varying according to the mode and the specific performance.
 
\begin{figure}
 \centerline{
 \includegraphics[width=0.9\linewidth]{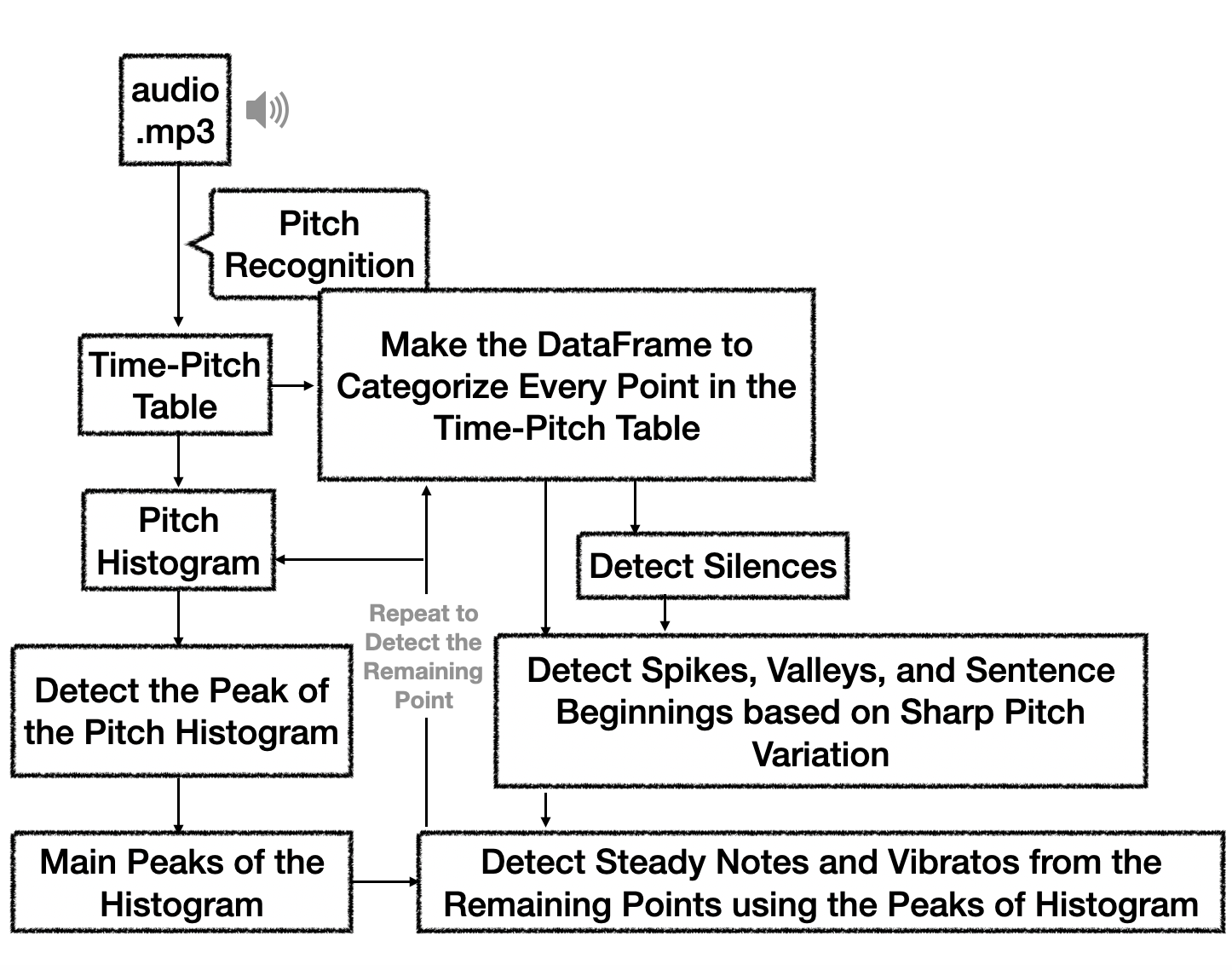}}
 \caption{Time-Frequency Data Processing}
 \label{algorithm}
\end{figure}

The process described in Sections \ref{pitch recog} to \ref{main notes} is summarized in Figure \ref{algorithm}, which outlines the sequential steps from pitch recognition to the identification of notes and various ornamentations.

\subsection{Pitch Recognition} \label{pitch recog}

There are various algorithms used for pitch recognition, all of which rely on the fundamental frequency of sound to measure pitch. The fundamental frequency is the lowest frequency in the sound wave and corresponds to the most prominent perceived pitch. Monophonic voice pitch estimation algorithms have been widely studied in Music Information Retrieval (MIR) literature. Gomez et al. have assessed the CREPE  \cite {KSLP-7} and pYIN \cite{MD-8} algorithms as state-of-the-art methods for recognizing pitch in monophonic voice  \cite {GBBC-9}. Their evaluation on the iKala dataset showed similar results for both algorithms, with pYIN achieving 91\% Raw Pitch Accuracy and CREPE 90.5\%. Another method for monophonic pitch recognition is SPICE  \cite {BGSpice9-a}, which uses a self-supervised learning approach. For this study, we selected pYIN and corrected octave errors when needed. Pitch-frequency tracks were generated from the audio recordings using the pYIN method through Sonic Annotator and exported as \emph{.csv} files. These files were then supplied to the transcription program. Because the subsequent transcription stages operate on pitch data in \emph{.csv} format, pYIN could be replaced with another pitch-estimation method in future experiments.

We generated the pitch-frequency tracks using Sonic Annotator with the following parameters: 
step size: of 256
block size: 2048
low amplitude suppression: 0.1
onset sensitivity: 0.7
pruning threshold: 0.1
threshold distribution: 2. 
We converted the sound frequency (in Hz) to the cents system, where each octave equals 1200 cents. The formula below calculates the interval between two notes with frequencies $f_1$ and $f_2$, measured in cents.

\begin{equation}
1200 \times \log_2( \frac{f_2}{f_1})
\end{equation}

After extracting the pitch data from the audio file, we construct a dataframe—a tabular data structure commonly used in Python programming for organizing and processing data—where the number of rows corresponds to the number of points in the pitch data table. This dataframe contains columns that categorize the points into various groups: solid notes, silence, spikes, valleys, sentence beginnings, and uncategorized. Each row in the dataframe is analyzed and assigned to one of these categories based on specific criteria.

\subsection {Tuning System and Histogram}

\begin{figure}
 \centerline{
 \includegraphics[width=0.9\linewidth]{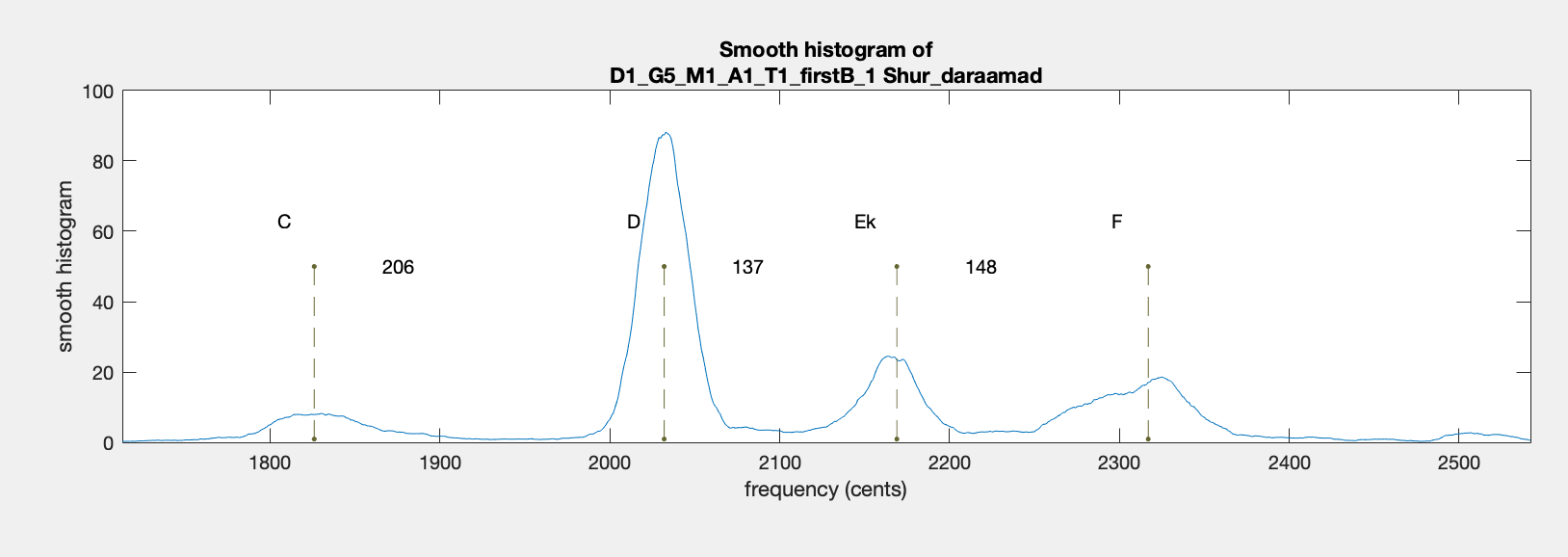}}
 \caption{Pitch Histogram for Example 1}
 \label{fig:Pitch histo}
\end{figure}

At this stage, it’s necessary to determine the tuning system of the performed piece. Given that tunings in microtonal music are fluid and often contain subtle variations, we use a pitch histogram to identify the primary pitches employed in the performance. To create the histogram, we calculate the total occurrences of each fundamental frequency. Figure \ref{fig:Pitch histo} presents the pitch histogram for Example 1, where the vertical axis represents the proportional total duration of each frequency (in cents).  In order to find the main peaks of the histogram we ignore the fluctuations and smooth the histogram using the moving average.

The pitch histogram is widely used in Music Information Retrieval (MIR) research. For instance, Bozkurt et al. employed it in their analysis of Turkish Makam music \cite{B-10}. In musical traditions where intervals and note frequencies are not fixed, pitch histograms prove valuable in identifying the frequencies of individual notes and the intervals between them in performed pieces. This technique has been thoroughly discussed for analyzing performances in Iranian music \cite{S-11}, \cite{S-12}, \cite{S-6}, \cite{BN-1}, and \cite{S-13}. Koduri et al. also utilized pitch histograms to examine intonation in Carnatic music \cite{K-14}.

\begin{figure}
 \centerline{
 \includegraphics[width=0.9\linewidth]{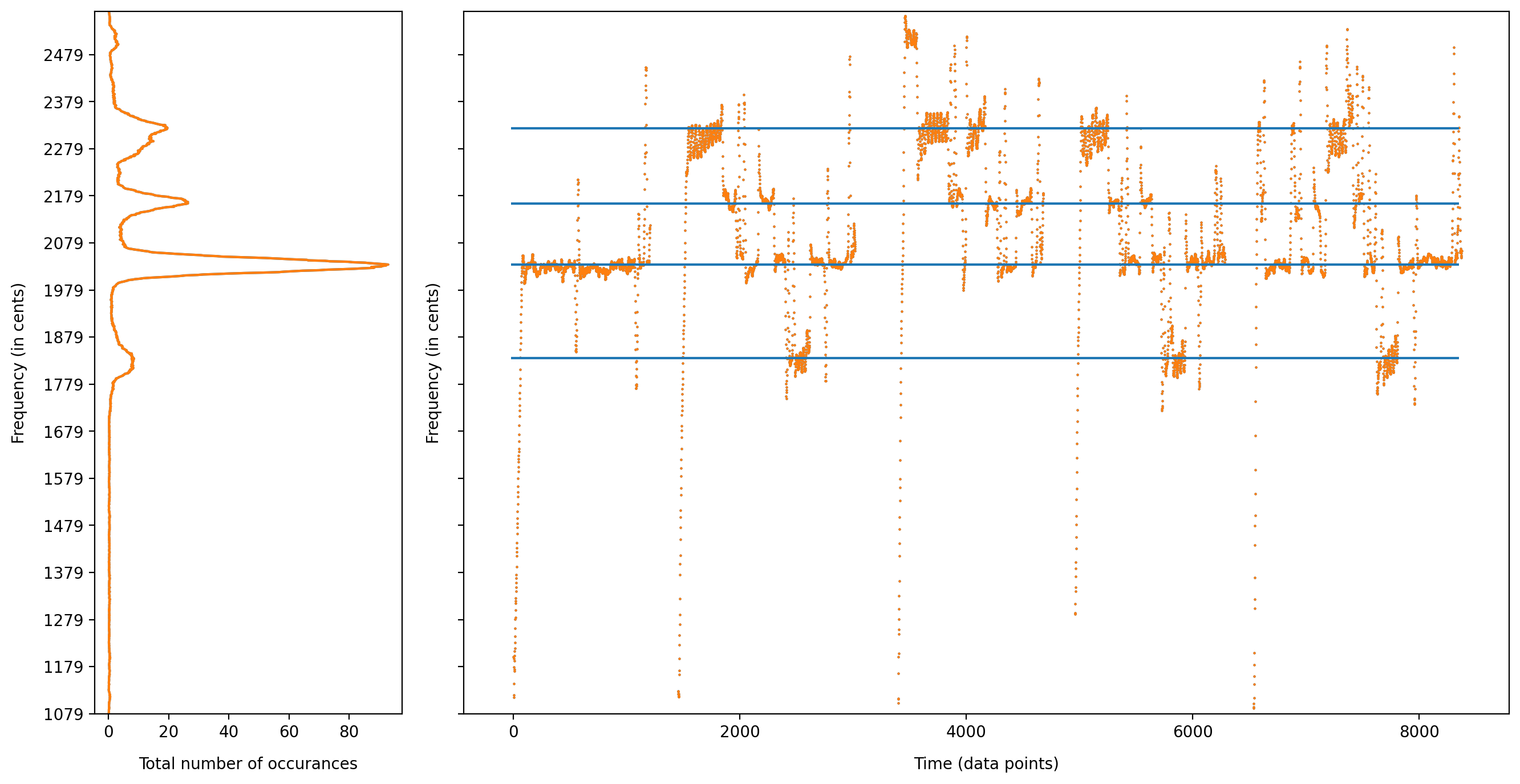}}
 \caption{Time-Frequency with Marked Peaks of the Histogram in Example 1}
 \label{histo-time}
\end{figure}

We identify the primary peaks in the pitch histogram, designating them as the main tones used in the piece. Users also have the option to specify the \emph{shāhed} note, enabling the transcription’s reference point to align precisely with their needs. Additionally, users can transpose the piece to begin from any chosen note. If the user does not specify particular \emph{shāhed} note, the code defaults to those commonly used in the most prevalent versions. Each blue horizontal line in Figure \ref{histo-time} represents a corresponding main peak (in the left subplot) of the histogram on the time-frequency graph for Example 1.

The automatic transcription captures the pitches exactly as they appear in the performance, preserving both their occurrence and precise intervals. In the final output, we visually distinguish the main structural notes from auxiliary or ornamental notes using different colors. Additionally, we have implemented a theory-informed transcription feature that filters notes based on theoretical expectations within a specific maqam (or gushe), catering to users familiar with the repertoire. This feature can be particularly useful for instrumentalists aiming to replicate the vocal performance faithfully while adhering to theoretical conventions.

\begin{figure}
 \centerline{
 \includegraphics[width=0.9\linewidth]{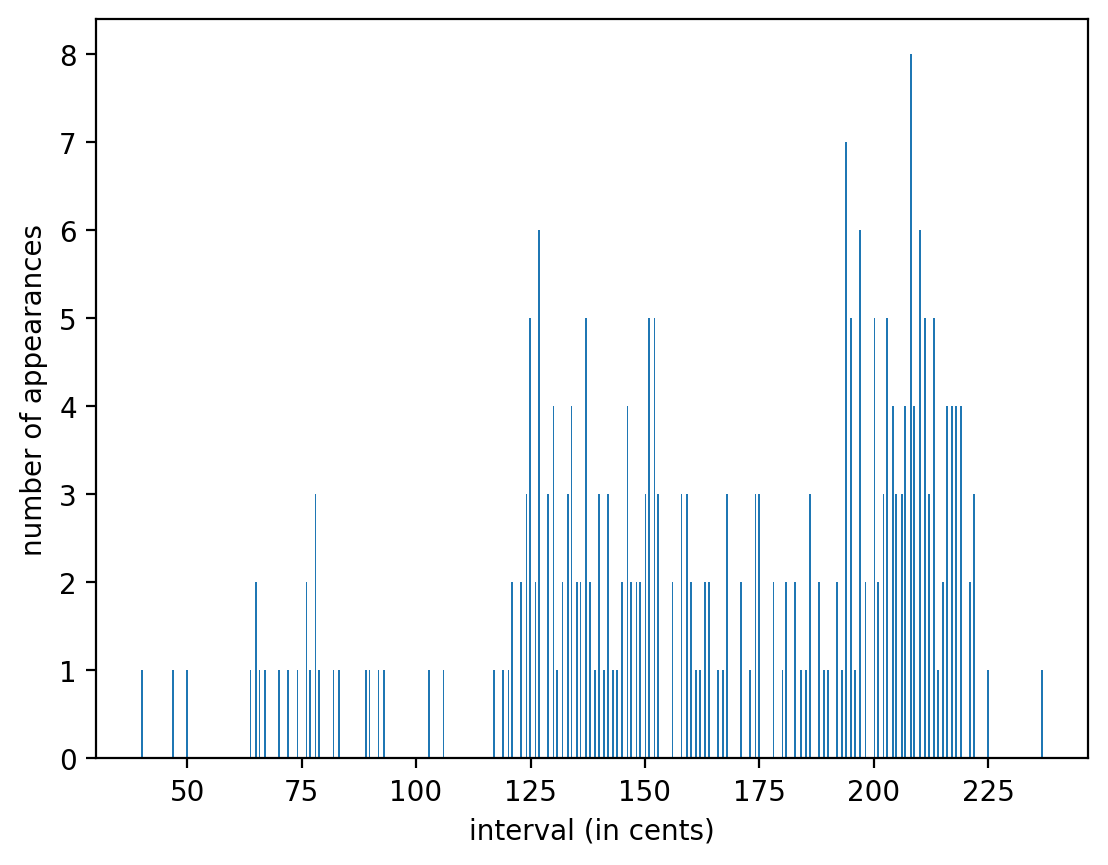}}
 \caption{Histogram of Second Intervals in the Scales of the First Five Dastgāhs in Karimi’s Radif.}
 \label{fluidity}
\end{figure}

Figure  \ref{fluidity} illustrates the fluidity of intervallic structures in Iranian classical music, a key factor contributing to the complexity of automatic transcription. This figure presents a histogram of all second intervals occurring in the scales of the first five dastgahs in Karimi’s Radif: Shur (15 pieces), Abuatā (9 pieces), Bayāt-e Zand (12 pieces), Afshāri (9 pieces), and Dashti (9 pieces).

\subsection {Phrase Onset and Expressive Contours of Tahrir}

After identifying the main peaks in the histogram, we then detect all spikes and sharp valleys in the time-frequency domain by analyzing parameters related to curve slopes. These features arise in two distinct musical contexts. First, at the beginning of each musical phrase, where the vocalist starts from a low pitch and quickly ascends to the target pitch. Second, they appear in all \emph{tekyes} and \emph{eshares}, which are elements of \emph{tahrir} ornamentation \cite{S-12},\cite{Mir-1}. 

\begin{figure}
 \centerline{
 \includegraphics[width=0.9\linewidth]{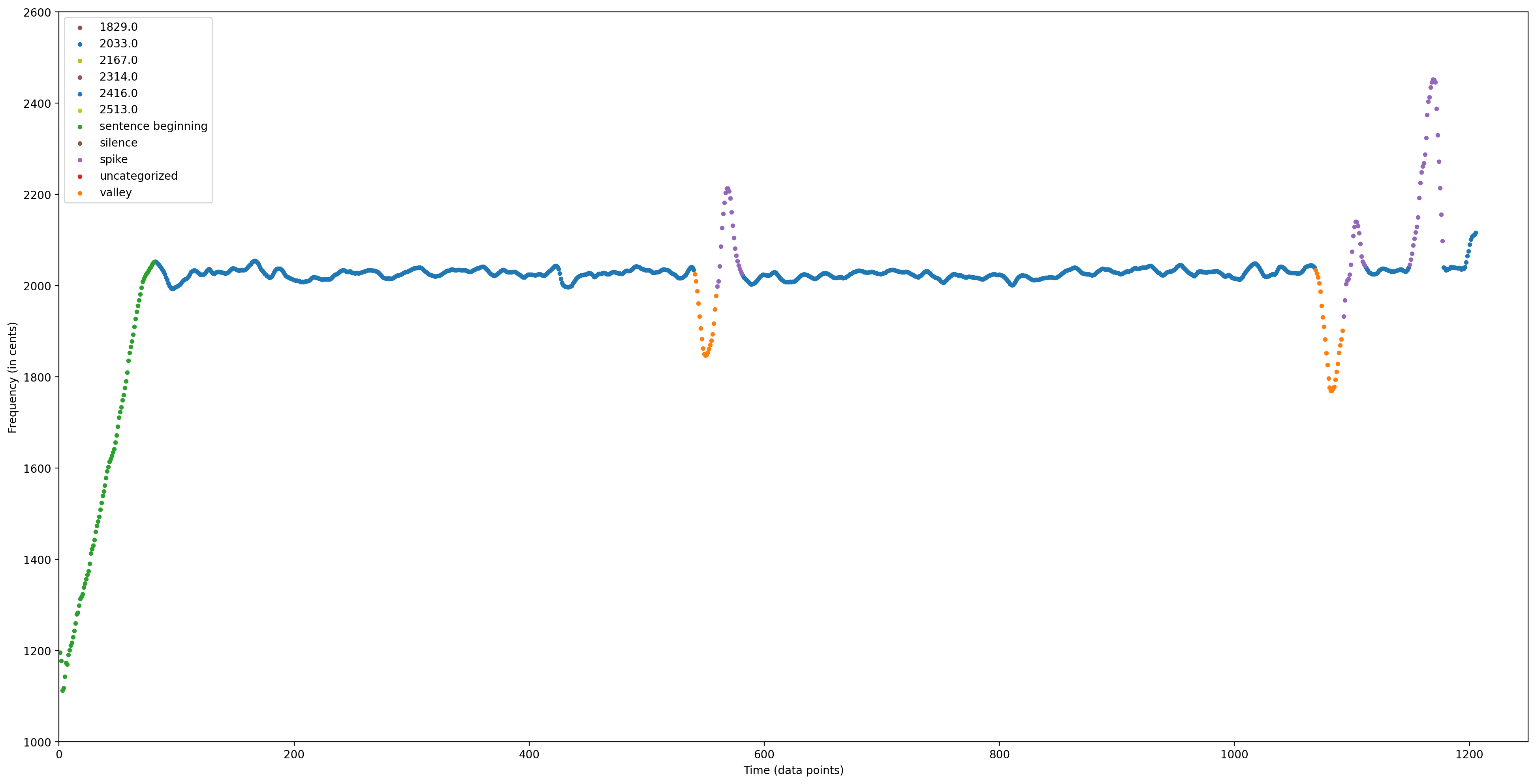}}
 \caption{The Time-Frequency Graph of the First Frame of Example 1 with Marked Notes and Ornamentations}
 \label{points}
\end{figure}

It is important to note that the specific intervals for these upward or downward sharp movements in pitch are approximate, as vocalists do not aim for exact pitches but rather perform these as expressive musical gestures. For this reason, transcribing these gestures as exact notes and failing to distinguish them from the main pitches can lead to a loss of essential information in the transcription of Iranian classical vocal music. On fretted or plucked instruments, instrumentalists typically play one degree above or below within the musical maqam to emulate the upward or downward movements found in vocal ornamentation. Figure \ref{points} illustrates the beginnings of sentences, spikes, and valleys for the first frame of Example 1.

\subsection {Main Notes and Secondary Notes} \label{main notes}

To identify the main notes, we search for frequencies within a defined neighborhood around the peaks of the pitch histogram. Once we locate clusters of consecutive points that belong to the same note, we examine the gaps between these clusters using a higher threshold to determine if they should be connected as part of the same note.

Pitch identification proceeds iteratively. The initial histogram identifies the principal stable pitches, while subsequent histograms are calculated from residual unassigned samples to recover shorter or less frequent secondary notes. After each pass, newly identified samples are assigned to their nearest eligible pitch category before the remaining data are reconsidered. These secondary notes may be very brief in duration or performed unintentionally by the vocalist. They typically fall outside the theoretical range of the maqām or dastgāh being performed and therefore require a distinct analytical approach. 

Figure \ref{points} illustrates the notes identified in the first frame of Example 1, each marked with its corresponding frequency in cents. These notes are distinguished based on the peaks of the pitch histogram, which represent the most stable and frequently occurring pitches in the performance, forming the tonal framework of the transcription.

\subsection {Tempo}

The music analyzed in this study is free-rhythm, presenting the challenge of notating the duration of each note once their boundaries have been established in the audio. To avoid overly complex transcriptions, we decided not to include note durations shorter than a 32nd note. To estimate a notational reference tempo for the piece, we define a broad search range for the duration of a quarter note, from 0.1 to 2 seconds (equivalent to nominal musical tempi of 600 to 30 beats per minute, respectively). We then incrementally compute (using a step size of 0.001) the cumulative error across the entire piece for each potential tempo. To prevent smaller values for the quarter note from automatically yielding smaller errors, we apply a weight to balance this effect.

\begin{figure}
 \centerline{
 \includegraphics[width=0.9\linewidth]{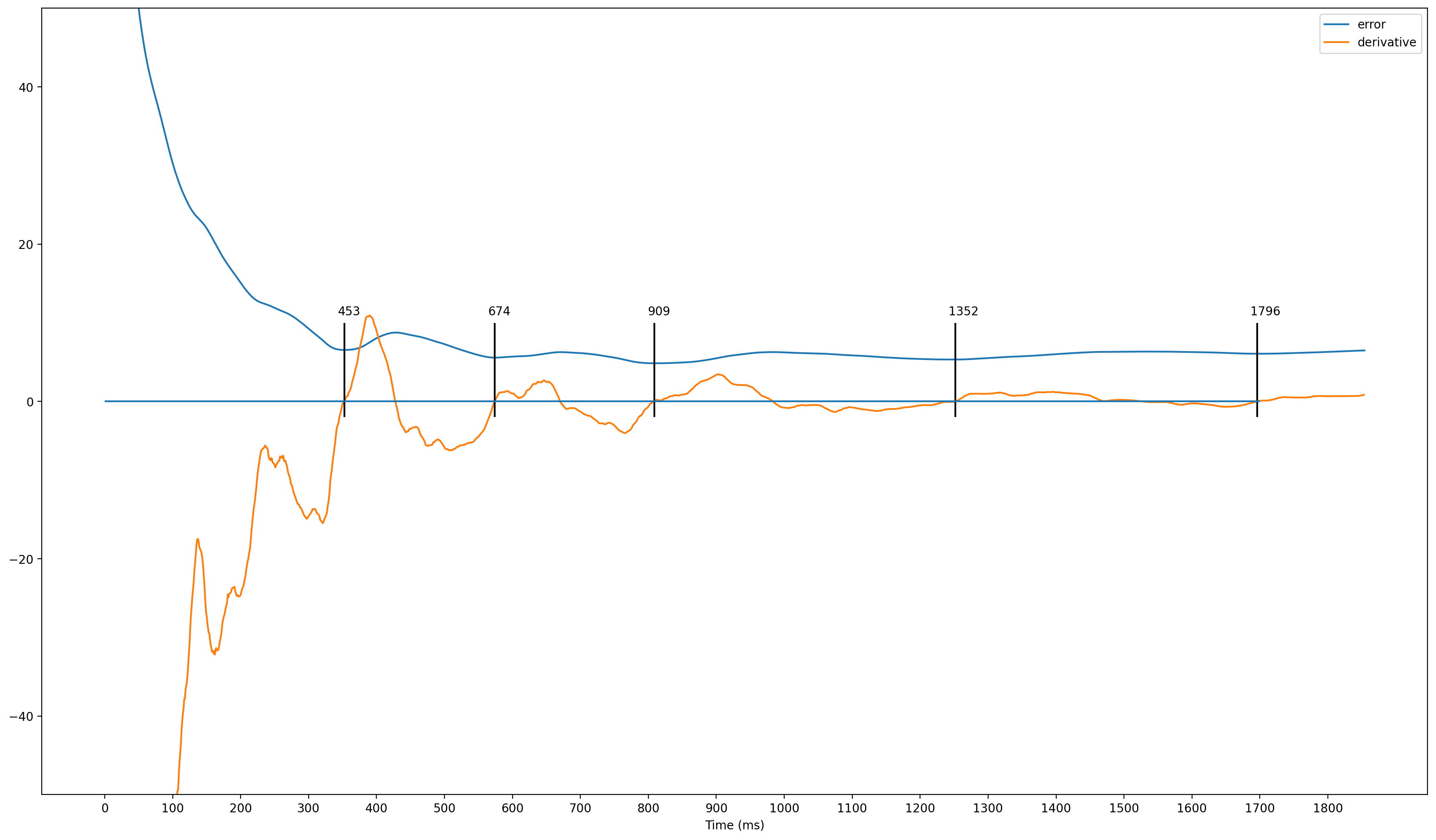}}
 \caption{Error and Derivative in Finding the Tempo for Example 1}
 \label{tempo}
\end{figure}

This process produces a curve with multiple local minima. We select the absolute minimum of this curve as the estimated tempo. Figure \ref{tempo} illustrates this process for Example 1: the blue curve represents the weighted error, and the orange curve shows its derivative. The vertical black lines mark the minima, which correspond to points where the derivative changes sign from negative to positive. The absolute minimum in this case is at 909 \emph{ms} which give the tempo 66.

The ratio between consecutive local minima further validates the method. Given that we expect the main durations to be 1, 1.5, and 2 times the base duration—yielding theoretical ratios of 1.5 and 1.33—the ratios obtained from the graph in Example 1 are 1.49, 1.35, and 1.33, which align closely with these theoretical expectations.

All the local minimums = 
$$\left[ 453 \ 674 \ 909 \ 1352  \ 1796 \right]$$

Ratio between the local minimums = 
$$\left[1.49 \ 1.35 \ 1.49 \ 1.33 \right]$$

\section{From Musical Information Table to Transcription and MIDI}\label{Transcription Table}

At this stage, we compile a table containing all the extracted information from the audio, including the beginnings of sentences, solid notes, and ornamentations along with their respective durations. This table is then converted into a text-based notation system using TinyNotation syntax, allowing integration with the \emph{music21} library. Once the transcription is generated as a .musicxml file using \emph{music21}, we apply pitch bends to capture the neutral intervals characteristic of Iranian music and produce a corresponding MIDI file. 

\begin{figure}
 \centerline{
 \includegraphics[width=0.9\linewidth]{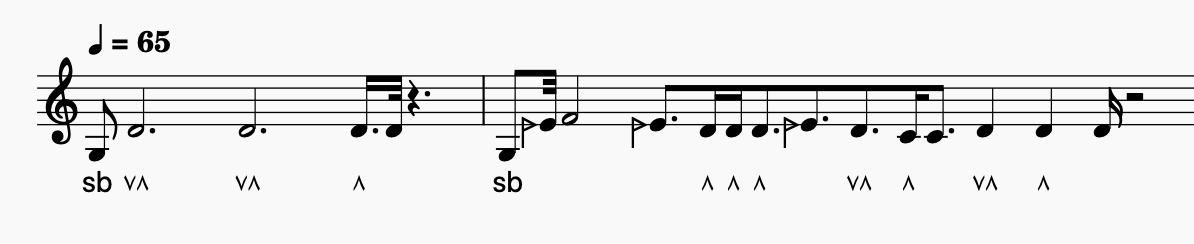}}
 \caption{Generated Transcription of the First Two Sentences of Example 1}
 \label{finalnote}
\end{figure}

Figure \ref{finalnote} presents the final transcription generated by our code for the first two sentences of Example 1. Similar to Masoudieh's transcription, the p-shaped sign for E indicates that E is half-flat. Additionally, we have introduced three new symbols to better represent the nuances of the vocal performance. The first symbol (sb) marks the beginning of a sentence, where the vocalist ascends rapidly from an indeterminate low pitch to the initial notated pitch. In accompanied performances, an instrumentalist may represent this gesture using a pitch approximately a fifth, an octave, or an octave plus a fifth below the initial pitch; these intervals describe an accompaniment convention and do not imply that the vocalist begins on that exact pitch. The second and third symbols, upward and downward arrowheads, denote sudden pitch shifts in each direction in tahrir, where the precise frequency is less significant than the expressive contour of the ornamentation.

\section{Visualizing and Analyzing Transcription Results}

After generating the music transcription (\emph{.MusicXML}) and MIDI file, we construct a time-series representation of the MIDI that replicates the subtleties of various culture-specific ornamentations. We then apply Dynamic Time Warping (DTW) to align the audio with this time-series, displaying the results on a single graph. This visualization supports detailed analysis by musicians and musicologists. Each note on the graph is annotated with its name and duration. For note naming, we follow the \emph{music21} convention: uppercase letters for lower octaves, lowercase letters for upper octaves, and lowercase letters with an apostrophe for even higher octaves.

\begin{figure}
 \centerline{
 \includegraphics[width=0.9\linewidth]{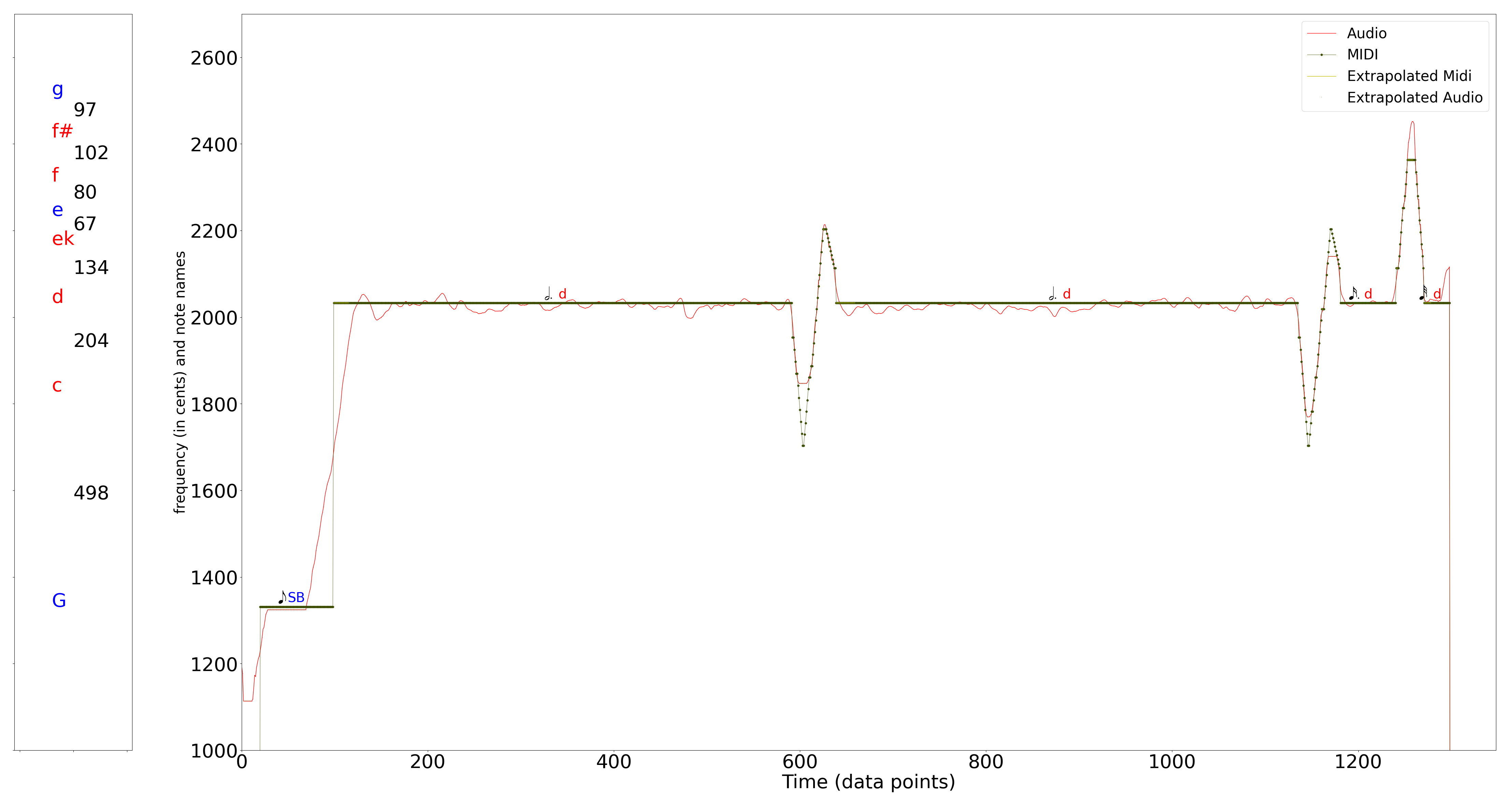}}
 \caption{Visualizing the Pitch and Generated Transcription, Example 1, Sentence 1}
 \label{finalsent1}
\end{figure}

\begin{figure}
 \centerline{
 \includegraphics[width=0.9\linewidth]{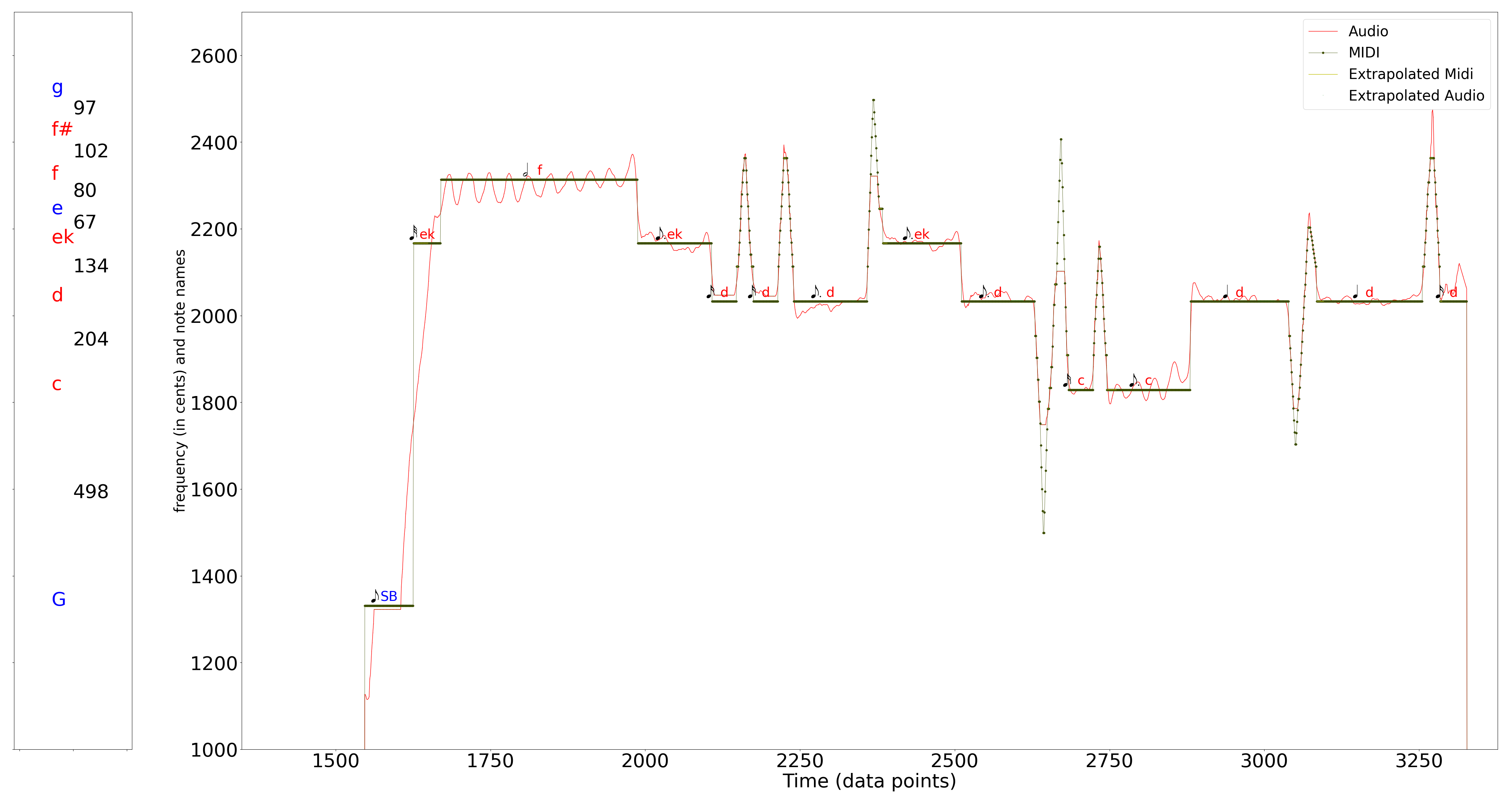}}
 \caption{Visualizing the Pitch and Generated Transcription, Example 1, Sentence 2}
 \label{finalsent2}
\end{figure}

On the left side of the graph, a bar displays all the notes used in the transcription along with their corresponding intervals. To enhance the analysis, we use distinct colors to differentiate between main notes and secondary notes. Additionally, spikes and valleys corresponding to \emph{tekyes} and sentence beginnings are marked separately on the graph. Figures \ref{finalsent1} and \ref{finalsent2} illustrate the final sequence matching results for Sentences 1 and 2 of Example 1, respectively.

\subsection{Expert-in-the-Loop Transcription Correction}

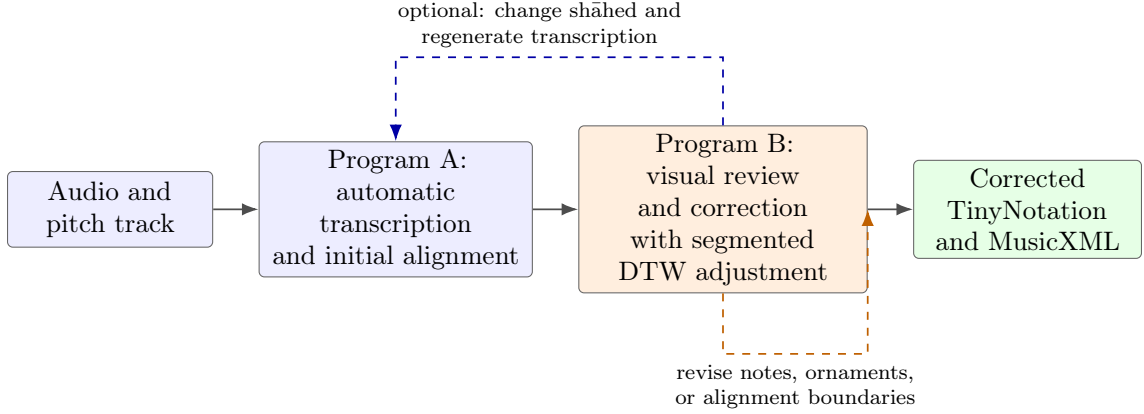
\begin{figure}[t]
\centering
\begin{tikzpicture}[
  font=\small,
  node distance=7mm and 6mm,
  process/.style={draw=black!60, rounded corners=2pt, fill=blue!7,
    align=center, minimum height=10mm, text width=25mm, inner sep=3pt},
  expert/.style={draw=black!60, rounded corners=2pt, fill=orange!13,
    align=center, minimum height=10mm, text width=28mm, inner sep=3pt},
  output/.style={draw=black!60, rounded corners=2pt, fill=green!10,
    align=center, minimum height=10mm, text width=28mm, inner sep=3pt},
  flow/.style={-{Latex[length=2.2mm]}, line width=0.65pt, draw=black!70},
  feedback/.style={-{Latex[length=2.2mm]}, line width=0.65pt,
    draw=orange!75!black, dashed},
  rerun/.style={-{Latex[length=2.2mm]}, line width=0.65pt,
    draw=blue!65!black, dashed}
]
\node[process] (audio) {Audio and\\pitch track};
\node[process, right=of audio, text width=34mm] (automatic)
  {Program A:\\automatic transcription\\and initial alignment};
\node[expert, right=of automatic, text width=36mm] (review)
  {Program B:\\visual review and correction\\with segmented DTW adjustment};
\node[output, right=of review] (export) {Corrected\\TinyNotation and MusicXML};

\draw[flow] (audio) -- (automatic);
\draw[flow] (automatic) -- (review);
\draw[flow] (review) -- (export);
\draw[feedback] (review.south) -- ++(0,-8mm) -|
  node[pos=0.25, below, font=\scriptsize, align=center]
  {revise notes, ornaments,\\or alignment boundaries} (review.east);
\draw[rerun] (review.north) -- ++(0,9mm) -|
  node[pos=0.28, above, font=\scriptsize, align=center]
  {optional: change shāhed and\\regenerate transcription} (automatic.north);
\end{tikzpicture}
\caption{Expert-in-the-loop correction framework. Program A generates the initial transcription and audio--MIDI alignment. Program B supports revision of notes, ornament markers, and alignment boundaries, including segmented DTW recomputation, before corrected symbolic files are exported. Program A is revisited only when a change to the shāhed requires the transcription to be regenerated.}
\label{fig:expert-loop}
\end{figure}

Figure~\ref{fig:expert-loop} summarizes the expert-in-the-loop correction workflow, which uses two separate programs. The automatic transcriber (Program A), described in the preceding sections, generates the symbolic transcription and a graph combining the audio pitch track with the DTW-aligned MIDI representation. The companion visual editor (Program B) is a separate program developed to support review and correction of that output. It displays the audio pitch track and symbolic transcription within the same time-aligned view. Musicians and music engravers can inspect individual passages, listen to selected segments, and modify note pitches and durations. For a selected passage, users can listen separately to the original recording, the extracted pitch track, and the synthesized MIDI representation, facilitating direct comparison between the performance and its symbolic representation. The interface also supports the insertion or removal of notes and ornament markers. Users can define or adjust audio--MIDI correspondence boundaries, after which DTW is recomputed within the resulting segments while preserving previously established boundaries. Corrections can be saved as editing sessions and exported as updated TinyNotation and MusicXML files.

Program A was applied to 165 pieces: 145 pieces from Karimi's radif in the IRMA dataset \cite{IRMA2025} and 20 additional recordings of Karimi made by Tsuge \cite{Tsuge1974}. Its automatically generated transcriptions were inspected by the authors and received professional music-engraving review from Joel Rodriguez Caraballo. Program B was tested separately on several pieces to verify its review, editing, alignment-adjustment, and export functions. The editor is designed to enable musicians and music engravers to correct automatically generated transcriptions without directly modifying the underlying code. Together, the two programs preserve the efficiency of automatic transcription while allowing culturally and musically informed decisions to remain under expert control. This is particularly important for Iranian classical music, where oral transmission, flexible timing, expressive microtonal variation, and ornamentation may be oversimplified by a fully automatic representation. The editor therefore treats computational transcription as an assistive and revisable interpretation rather than an authoritative replacement for musical expertise.

\section{Conclusion}

This study presented a computational workflow for transcribing microtonal, free-rhythm vocal music, using Iranian classical music as a case study. The method combines pitch recognition, multi-level pitch-histogram analysis, rule-based identification of structural and ornamental events, tempo approximation, symbolic representation with \emph{music21}, and DTW-based alignment between the audio and the generated transcription. The resulting notation distinguishes structural notes from expressive gestures such as sentence beginnings, \emph{tekye}, and other forms of \emph{tahrir}, while retaining the microtonal and temporal characteristics of the performance.

Application of the automatic transcriber to 165 pieces from Karimi's radif demonstrated that it can be used across a substantial body of repertoire rather than only for the worked example presented in this paper. The companion visual editor, tested separately on several pieces, supports inspection of the aligned audio and notation, revision of pitches, durations, ornament markers, and alignment boundaries, and export of corrected TinyNotation and MusicXML files. The method therefore provides an interpretable starting point for large-scale transcription while preserving a central role for expert musical judgment.

From a cultural-heritage perspective, the workflow supports the documentation of performance characteristics that conventional notation may obscure, including flexible microtonal intervals, free-rhythm timing, and culturally specific ornamentation. The audio--transcription alignment keeps the symbolic representation traceable to the original recording, while expert correction helps prevent automatically generated notation from being treated as an authoritative replacement for the performance. The resulting materials can support archival access, comparative analysis, pedagogy, and future computational research while preserving the recording as the primary musical document.

The present work does not treat an automatically generated score as a definitive representation of a performance. The selection of thresholds, interpretation of expressive gestures, and comparison with reference transcriptions remain sensitive to musical context. Future work will include a formal quantitative evaluation of transcription accuracy, analysis of agreement among expert reviewers, and testing on performers and related modal traditions beyond Karimi's radif. These developments will help determine which stages can be further automated without obscuring the culturally specific features that the transcription is intended to preserve.

\appendix
\section{Glossary of Iranian Classical Music Terms}
\label{names}

Below is a list of all terms and concepts related to Iranian classical music, along with their definitions, that are referenced throughout this study \cite{S-6}.

\underline{ \emph{Darāmad}}: The opening gushe of each dastgāh, typically performed first, introduces the dastgāh by establishing its primary modal framework. This gushe is characterized by its lack of meter.

\underline{ \emph{Dastgāh}}: The radif is composed of seven dastgāhs and five āvāzes (secondary dastgāhs). Each dastgāh includes multiple pieces (gushes) organized in a structured sequence. This level of organization arranges gushes in relation to each other, guiding radif students through modulation pathways.

\underline{ \emph{Eshāre}}: (literally hint) A grace note following the main note with a lower or higher pitch. It can also be a sequence of secondary notes featuring both higher and lower pitches and is an aspect of tahrir \cite{T-33}.

\underline{ \emph{Gushe}}: (literally corner) Refers to the individual units that make up each dastgāh or āvāz. Each gushe has unique modal, melodic, or rhythmic qualities and is organized in a particular sequence within a dastgāh.

\underline{ \emph{Ist}}: (literally stand) Denotes the ending note of an intermediate phrase distinct from the shāhed \cite{W-15}.

\underline{\emph{Maqām}}: In contemporary Iranian musicology, the term maqām is often used without a precise definition, although some sources do define it. In this work, we use Alizāde’s definition: maqām, or mode, is a scale in which certain tones play particular roles, and specific melodic motifs may be characteristic of pieces in that maqām. For instance, in a scale like [C D E F G A Bb C], a maqām is established if we emphasize G and pause on D. Alizāde further explains that recurring melodic models might appear across pieces within the same maqām, marking it as a defining feature. Additionally, Iranian music may employ tetrachords or a pair of tetrachords rather than a complete scale to define maqām \cite{Al-1}.

\underline{\emph{Moteghayyer}}: (literally alterable or variable) Describes a pitch that may be replaced by another, either a quartertone or semitone higher or lower, within the gushe \cite{W-15}.

\underline{\emph{Radif}}: (literally series or row) The repertoire or framework of Iranian classical music, consisting of hundreds of primarily free-rhythmic pieces (gushes) arranged into 12 or 13 systems called dastgāhs. Musicians internalize the radif through memorization and repetition, using it as a basis for improvisation.

\underline{ \emph{Shāhed}}: (literally witness) The most prominent pitch in a gushe, primarily distinguished by its relatively longer duration \cite{W-15}.

\underline{\emph{Tahrir}}: A unique vocal technique, similar to yodeling, involving sudden shifts in pitch and characterized by various styles.

\underline{\emph{Tekye}}: (literally leaning and emphasize) A rapid transition in tahrir from the main note to a higher-pitched note, commonly known as tekye. It is a grace note following the main note, with a pitch typically one to four semitones above the main note.

\section*{Code Availability}

Source code and release information for the transcription and alignment system and its companion visual editor are being prepared for public release and will be made available through the project repository at \url{https://github.com/SepiSha/microtonal-music-autotranscriber}.

\bibliographystyle{plainnat}
\bibliography{references}

\end{document}